\documentclass[11pt,twoside]{article}
\usepackage{asp2004}
\usepackage{epsf} 
\usepackage{graphics}

\usepackage{lscape}
\def\edcomment#1{\iffalse\marginpar{\raggedright\sl#1\/}\else\relax\fi}
\reversemarginpar

\begin{document}
\title{Astronomical databases, Space photometry and time series analysis:
Open questions}
 \author{L. Eyer}
\affil{Observatoire de Gen\`eve, CH-1290 Sauverny, Switzerland}

\begin{abstract}
How to analyze TeraBytes of photometric data, and extract knowledge on variable stars? How to detect variable phenomena? How to combine different photometric bands? Which algorithm to search for periods?  How to characterize and classify the detected variable objects? Many questions, but certainly no definitive answers yet. We present several aspects which are at the interface of photometric surveys and variable stars. Fully automated analyses of photometric surveys are still not at an optimized level. We will take the example of a future survey, the Gaia mission project of the European Space Agency, to show different steps of a possible automated pipeline scheme. Principal component analysis can be applied to the Gaia photometric bands. We give some illustrative examples of classification methods such as Support Vector Machine, Self-Organizing Map, or Bayesian classifier.

\end{abstract}

\vspace{-0.5cm}
\section{General Introduction}
In this presentation, we will focus on photometric surveys and the variability analysis they require to extract information on variable stars\footnote{Other questions may be raised for radial velocity, astrometric surveys, and, the gathering of data of different natures.}.

A survey is characterized by its surveyed luminosity interval, its photometric precision, its sky coverage and its sampling strategy. The survey impact on variable stars can vary a lot depending on these four parameters. 

Indeed, we remark that the results of different surveys have different returns on variable stars.
In Hipparcos, still the only modern whole sky survey with an associated systematic variability analysis (more than 10 years after its completion), nearly 10\% of the sources are variable. The observed stars by Hipparcos were coming from a survey and a selected sample. We remark that the fraction of variable stars in the Hipparcos survey is similar than in its selected sample, so that the high number of variable stars is not coming from stars specifically introduced in the selected sample because of their variability. 
ASAS (All Sky Automated Survey) has nearly 3\% of its sources variable, {\sc OGLE-II} nearly 1\%. Why do we have such differences between these surveys?

The variable star types and their numbers detected by surveys depend on the above mentioned considerations, i.e. on surveyed targets,  the interval of magnitude observed and the scanned sky region, the observational strategy, the overall instrument precision and also the associated variability analysis.

The complexity of the analysis is coming from various reasons:
the large number of sources, the diverse intrinsic natures of the observed stars (there is a real zoo of variable stars), the complexity of the flux or magnitude data
(because perfect data do not exist, there are outliers, the number of measurements may vary from one star to an other for the same survey, the noise depends on the magnitude), and the irregularity of the time sampling.

There are many surveys, which have been set-up for different reasons. There are surveys searching for microlensing, gamma ray bursts, planetary transits, Near Earth Objects, etc\ldots The field of photometric surveys has now been booming for one decade and continues to develop. For instance, we could mention some future surveys, like Pan-STARRS, LSST, and space missions like COROT, and Kepler, they will strongly impact the knowledge of variable stars.

There is one survey that we haven't mentioned yet, it is the Gaia mission, project of the European Space Agency (ESA). We will take this survey as an example and see what types of questions are being tackled for the preparation of some aspects of the data analysis.

\section{The Gaia mission}

Gaia is a cornerstone mission of ESA with a target launch date in 2011. Gaia will survey the stellar content of our Galaxy by measuring the position of stars and their dependence with time over 5 years, leading to the star parallax and proper motion. These astrometric measurements will be complemented by a spectrometer which will lead to the radial velocities and by two photometric systems which will allow the determination of stellar parameters such as the temperature, surface gravity and metallicity.

The astrometric and photometric measurements will be obtained for 1 billion stars down to $V=20$. The radial velocities will be obtained down to $V=17$.
The data size to be downloaded from the satellite position, around the second Sun-Earth Lagrangian point, to the Earth is estimated to 100-200 TeraBytes,
the whole database size is estimated to be 1-2 PetaBytes.

The data size, nature, diversity and inter-winding make the data processing of Gaia a challenge of the highest order.

For more detailed information on Gaia, consult the website:\\
\verb"http://www.rssd.esa.int/index.php?project=GAIA&page=index"

\section{How to detect variability?}
Different stars observed by Gaia will generally have different numbers of measurements. Different stars of different magnitudes will have different photometric errors. The photometric data of stars in a crowded area will suffer from perturbations that stars in an isolated environment won't have, etc\ldots The nature of the data of the Gaia survey and other surveys is quite heterogeneous.
In this context, the basic question is how to compare time series of different stars, how to detect variable stars on a similar basis? one answer can be to
use the hypothesis testing. We refer the reader to \cite{KS69} for a good introduction to hypothesis testing.

Hypothesis testing determines the probability distribution of a quantity, $t$,  under two hypotheses, $H_0$ the null hypothesis, $H_1$ the alternative hypothesis (though often the alternative hypothesis is difficult to model).
Typically $H_0$ is that the star is constant (only a measurement noise) and $H_1$ is variable.

To perform a test to accept or reject the $H_0$ hypothesis, we compute
the distribution of $t$ under $H_0$ and divide by a certain value of $t_{th}$, a threshold value, the possible values of $t$ into two regions, acceptance and rejection regions, with two associated probabilities. The probability of the rejection region is called the size of the test. Obviously this probability should be preferably small.

If we reject the hypothesis $H_0$ when the computed value of $t$ is in the rejection region, we have an estimation of the error of judgement we are doing if $H_0$ is true, we know its probability.
We can also define an other type of error, when a value of $t$ falls in the 
acceptance region under the $H_0$ hypothesis but $H_0$ is false. These two erroneous judgments are called type-I and type-II errors, or false negative in terms of $H_0$, and false positive in terms of $H_0$, respectively. For an example see the section~\ref{subs:trend} on trends.

We call p-value the probability that the variable $t$ has a value greater than the observed value of $t$ under the null hypothesis. In other words $H_0$ is rejected if the computed p-value is smaller than the size of the test, that is the observed value falls in the rejection region.

It is important first to fix the size of the test (at an error level that we would consider acceptable, for instance one per cent) and then perform the test. Otherwise our personal wishes may change the result of the test. 
~ \\

\noindent{\it Several tests to detect variability.}
The hypothesis testing and computation of p-values offer a way to compare different time series, and allow a judgement that a star is variable or constant (or non-detected as variable) on a same footing, that is the probability to announce that the star is variable when it is truly constant.

Two statistical tests can have different performances with respect to the alternative hypothesis $H_1$ when we have taken a given threshold for the type~I Error  (i.e. reject $H_0$ when it is true), see the example in section~\ref{subs:trend}
For these reasons, we are choosing several tests in the preliminary studies made for Gaia. These tests are the $\chi^2$ test, tests on the asymmetry and kurtosis
(they are related to three moments of the distribution), Abbe test \citep{JvN41},
a test to detect outliers, and a test on the presence of a slope. The three first tests are only using the magnitudes, the three last tests are taking into account the ordering of the data, or including the observation dates.

We should site an other way to detect variability. \cite{CKLE01}
detected directly the variability from a period search algorithms.
In order to establish a statistically good detection threshold, the method used is the permutation of the data (cf. section~\ref{subs:per}). 
~\\

\begin{figure*}[!ht]
\plotone{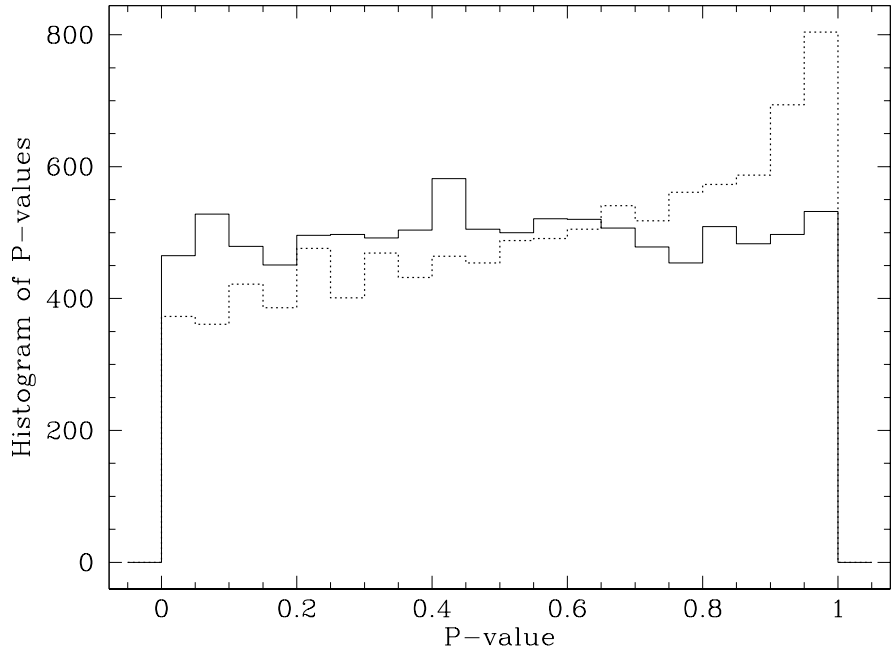}
\caption{Histogram of p-values of $\chi^2$ test. Solid line: correct estimation of noise; the distribution is flat. Dotted lines: noise slightly over estimated; the distribution is inclined.}
\label{fig:histopval}
\end{figure*}

\noindent
{\it Other use of p-values.}
Let us do a small digression on p-values.
The distribution of p-values if all the stars satisfy the $H_0$ hypothesis
should be uniform (flat). In usual cases, $\chi^2$ test is quite sensitive to variability, if a small fraction of the stars are variable, then the p-value of the test is very small, the histogram of p-values is flat except for a peak near zero.
In case the errors are indeed Gaussian, but are underestimated or overestimated, the histogram of the p-values will show it by not having a flat distribution. This method provides an easy way to see if the data is behaving as we expect it. We can also
use this property to estimate the noise level of the data, by adding a term in the error model in order to flatten the histogram of p-values. This was used by \cite{ADetal91} to estimate the instrumental noise of the {\sc coravel} instrument. We give in Figure~\ref{fig:histopval} an example of such procedure, we added an additional noise of 5 percent to the noise level used to compute the value of the $\chi^2$. We remark that the model with the correct noise estimation gives a flat distribution of its p-values (solid line). The model with slightly too large errors has an inclined distribution (dotted line).
~\\

\noindent
{\it Too many false detections?}
When we are dealing with large number of objects, the following problem arises:
If for instance we fix the threshold value even at a restrictive 1 per thousand and apply a test on the one billion stars of Gaia, if all the stars are constant, we will get 1 million stars declared variable though they are truly constant, this is too many. If we lower too much the size of the test, then we may miss interesting variable objects.

Fortunately with Gaia there are also two instruments (Astro-, Spectro-), and four sets of magnitudes: two full-light G-band (most precise, from Astro-Instrument) and GS (from Spectro), the Broad Band Photometric system (BBP, from Astro) and the Medium Band Photometric system (MBP from Spectro).
This will allow to have decent thresholds and control the problem of false detections. We mention the Pan-STARRS survey, its four telescopes will observe the same region of the sky (in one possible operational mode), this will also permit to diminish drastically the false detections.

If we want to be restrictive but do not have the possibility to rely on different independent measurements, we can select only the objects which are satisfying several statistical criteria. Some tests can have a certain level of independence and therefore we retain only very good candidates for variability. Apparently such studies of independence of tests are not so extensive and deserve a closer look. For example the $\chi^2$ test and the Kuiper test seem to have quite independent behaviors (Pourbaix, private communication).

\subsection{An example: Trend detection}
\label{subs:trend}
\begin{figure*}[!ht]
\plotone{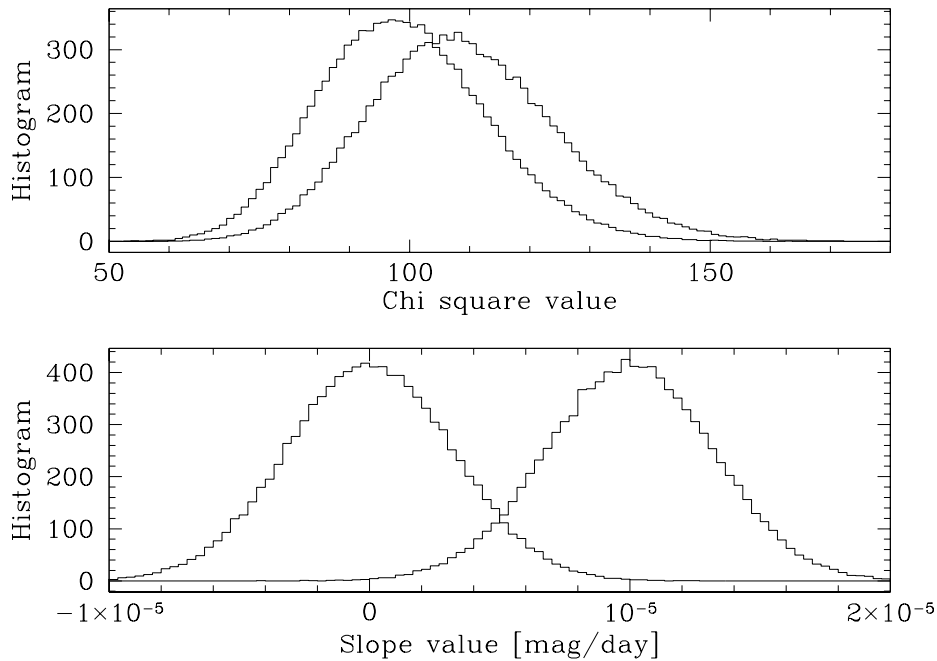}
\caption{Results of two variables, $\chi^2$ (top), slope of a trend model under two hypotheses, $H_0$ Gaussian noise, $H_1$ Gaussian noise a trend model.}
\label{fig:trends}
\end{figure*}
In this example, we illustrate several points of the previous sections. We show that the $\chi^2$ test is a poor way to detect trends. The power of the test is low, cf. Figure~\ref{fig:trends}.
 
We defined two variables, the $\chi^2$ (top figure) and the slope of a trend model through the data (bottom). For each of the two variables, we compute the distributions under two hypotheses: a) $H_0$, Gaussian noise (in both graphs the left distribution), b) $H_1$, a slope of $10^{-5}$ mag/day with same Gaussian noise as for $H_0$ (in both graphs the right distribution).

We remark that the two distributions of $\chi^2$ overlap, in such a way that
there are no good values of the $\chi^2$ which could serve as separating the distributions of the two different hypotheses. On contrary, the histograms of the slopes have only an overlap in the tails. To take a value of the slope between 0.3 and 0.6~$10^{-5}$ for the threshold would seem a good compromise for the false negatives and false positives.

\subsection{How to use efficiently a multi-color photometry?}
Gaia mission will get photometry in 21 bands. How to use efficiently a multi-color photometric system for the detection of variability? and how to perform the subsequent analysis? To make a variability detection in the most precise band may not be the most effective way to go ahead, because other bands can contain valuable variability information, even if they are less precise.

For example, some stars like the Ap stars, where the variability is due by the combination of the presence of surface spots and the rotation of the star on itself, can have very little detectable variability in a broad band filter, blocking effect and backwarming are happening in different spectral regions and may compensate each other on a broad filter. This is why in Hipparcos many known Ap stars were not detected as variable stars.

A simple idea, proposed by Paul Bartholdi, is to make a principal component analysis (PCA) on the whole set of bands. If $s_{njk}$ is the magnitude of the star $S_k$ in the filter $f_j$ at the time $t_n$, we may compute the covariant matrix for a certain star (we omit the indice k):

$$ C_{lm} = \frac{1}{N} \sum_n (s_{nl}-\bar{s}_l)(s_{nm}-\bar{s}_m), $$

where $N$ is the total number of measurements for the studied star,
$\bar{s}_l, \bar{s}_m$ are the mean magnitudes in the filters $l$ and $m$.
If the photometric system has $J$ filters, the matrix is $J \times J$.
The matrix diagonal terms are just the variance of the magnitude measurements in each filter of the system.

Then we can compute the eigenvalues of this matrix. The small and large eigenvalues are giving the following pieces of information:
1)~The smallest eigenvalue gives the minimum value of the noise after all correlated variations have been removed. So, it can be used to study the error budget.
2)~The ratio of the largest over smallest eigenvalues is a very sensitive indicator of the variability.
3)~The decomposition of the eigenvector with the largest eigenvalue can be used to search for periodicities. 
4)~The eigenvector with the largest eigenvalue can be used to describe the spectral energy distribution of the variation. Thus this information will be of great use for a classification scheme.

This method has been applied to a subset of standard stars of the Geneva photometry and it has been shown that the most stable stars have standard deviation of 2 milli magnitudes. For some stars previously declared constant, variability has been indeed detected from true stellar variability, from residual noise coming from atmospheric perturbations or reduction methods.

For the Gaia mission, we will test such procedures and will implement them.

\subsection{How to search for periodicity?}
\label{subs:per}
There are many period search algorithms. We try on the Gaia Variable Star Working Group webpage (under Tools section) to gather information on them, cf.
\verb"http://obswww.unige.ch/~eyer/VSWG". We can distinguish between different classes:
a)~Fourier type transform;
b)~Analysis of Variance (ANOVA), Phase Dispersion Minimization (PDM);
c)~String methods.
General studies have been done with simulations \citep{AHetal85}, and on a more analytical point of view \citep{ASC99} to compare different methods.

The question of making a test of hypothesis on the significance of the height
of peaks in a periodogram is not completely resolved yet. Though we know for many period search technics the statistics of the behavior of the heights at a given frequency (usually under the hypothesis of Gaussian errors), we do not know how to perform the test on a whole periodogram as we do not know how many independent frequencies we have, so we do not know how to fix correctly the threshold of false
negatives under $H_0$ (the approach of \cite{JHSB86} is generally not correct). This is not the case if the sampling is regular, the values of the power at the different frequencies are statistically independent.
There are several ways to circumvent this problem, one possible approach is to
perform simulations on random permutations of the time series. The advantage is that the permuted data sets have the same time points of observation, and the same noise level as the original data. Such an approach has been done in \cite{CKLE01}. An other way is a semi-analytical approach proposed by \cite{SP04} to determine the number of independent frequencies. There has also
been heuristic approaches to this problem as by \cite{RK97}.

In our opinion, there is still a large margin for improvement in this signal processing subject.
This is why for the preparation of the Gaia mission, we will launch a benchmark of period searches with simulations of the Gaia sampling on several variability types. Results will be communicated in due time.

\subsection{What features for characterizing the variability?}
If an object is detected as variable, the next question is how to describe it.
We would like to be able to separate the part of the signal which is due to an intrinsic property of the object and the part which is due to the measurement noise or any systematic external effects.

There are many possibilities: intrinsic variability (as an amplitude or a dispersion); moments of the magnitude distribution (skewness, kurtosis); time-scale(s), period(s); decomposition from PCA; for periodic signals the parameters of a Fourier decomposition can be used (however for stars having sharp features in their light curves, like eclipsing binaries, the Fourier decomposition encounters difficulties).

One recent solution proposed by Vasily Belokurov is to use an envelope of the power spectrum. Frequency bins are made and the highest power-peak within a given bin is used for representing the value of the power in this bin.

The features characterizing the variability are also linked to the classification methods which is used subsequently.

\subsection{How to classify variable objects?}

\begin{figure*}[!ht]
\plotone{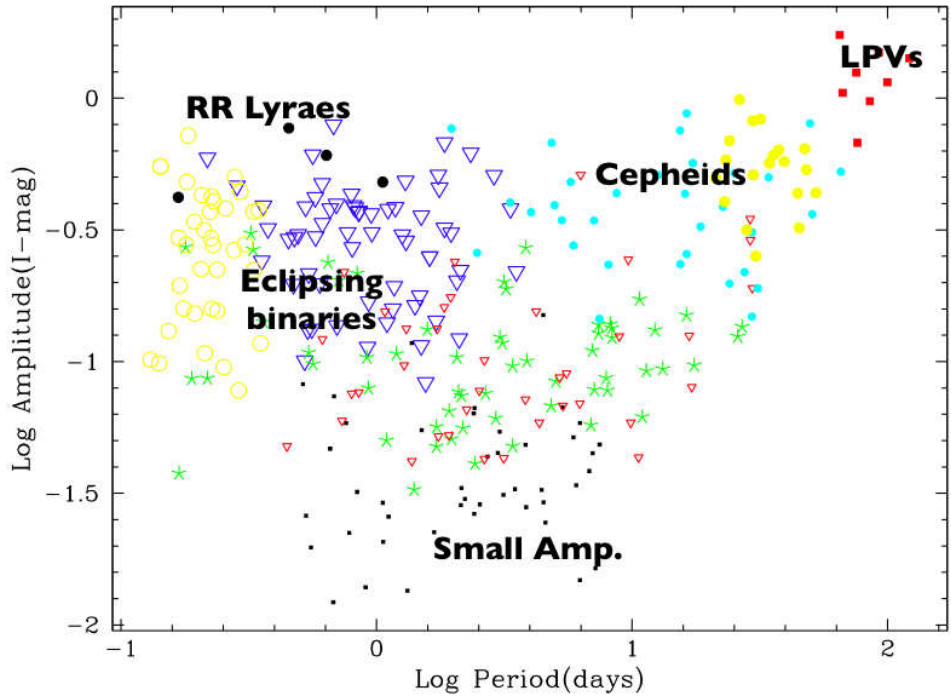}
\caption{Results of a Bayesian classifier on ASAS data.}
\label{fig:lecb}
\end{figure*}

The classification of a database is an other problem not yet resolved. There have been several approaches, some published others under developments. Previously {\sc ogle}, {\sc eros} or {\sc macho} groups have essentially extracted certain specific star groups, by designing some filters. {\sc Eros} survey is now developing a global approach (Marquette, private communication). We won't be exhaustive in the description of methods, but we show examples of some of them. What is important to retain is that this field 
is in its adolescence.  A benchmark of classification methods
is an other identified task that we will perform in the perspective of the Gaia mission. We have begun to familiarize ourselves with different methods:

\cite{LECB05} applied an unsupervised Bayesian classifier Autoclass on the All-Sky Automated Survey (ASAS) data \citep{GP00}. Unsupervised means that the algorithm is not fed with a previous classification. In principle it is therefore able to detect new variable star classes.  In total 1700 stars were classified. The used parameters to classify 302 "well" behaved periodic signals were the period, the amplitude, the phase difference and amplitude ratios derived from a Fourier series. The results are displayed in Figure~\ref{fig:lecb} in a $\log$(Period) $\log$(Amplitude) graph, such diagram
is a practical representation of the data. We see the results of the classification, with the different symbols. The classes found by this algorithm were easily identified with known classes of variable stars. The classification error level has been estimated to about~7\%. In these proceedings Jonas Debosscher is applying Autoclass to  Hipparcos data.

\begin{figure*}
\plotone{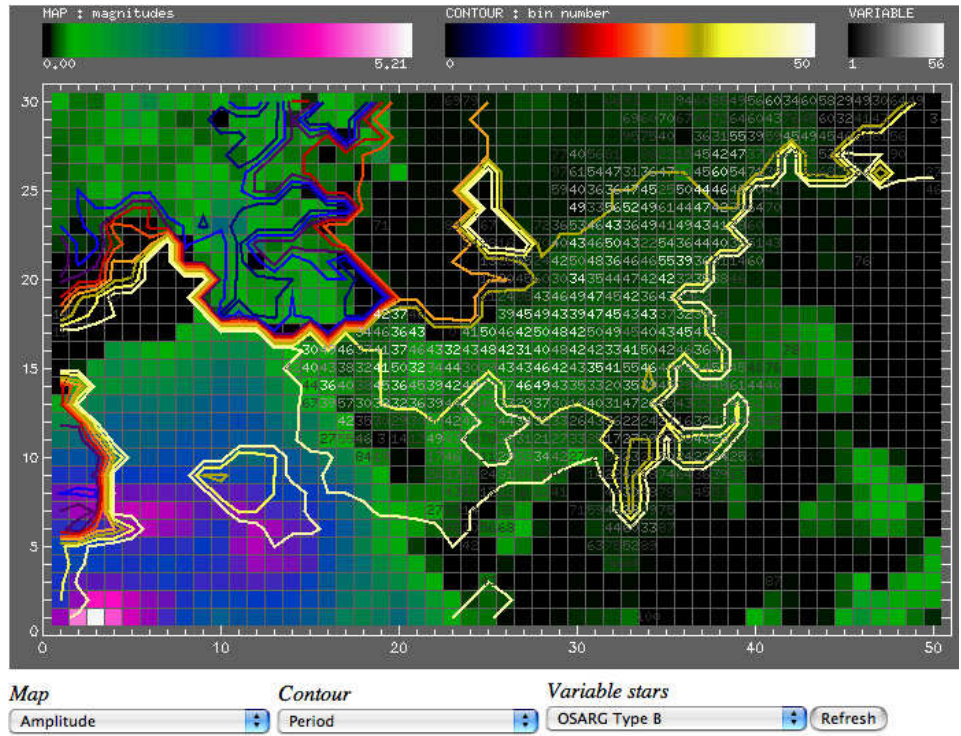}
\caption{Self Organizing Maps on {\sc ogle-II} bulge data. Colored map are the amplitudes, contour plots are the periods, the {\sc ogle} small amplitude variables (type B) are represented by the numbers. Courtesy of V.Belokurov}
\label{fig:som}
\end{figure*}

\cite{WEVB05} applied a self organizing map (SOM) on the 220,000 variable stars of {\sc ogle-II} bulge data \citep{PWetal02}. This method as applied by \cite{WEVB05} is unsupervised. The features are defined in the following way: a)~the maximum values of the power spectrum in 50 predefined bins are computed; b)~The 2nd and 98th flux percentiles are subtracted and converted into magnitude; c)~The ratio of the flux difference between the 98th percentile and the 50th (the median) over the flux difference between the 98th and 2nd percentiles is computed; d)~the $V-I$ color is extracted from the database. The total number of features is thus 53. The map is computed with a sub-sample of 60,000 stars. The result is displayed in a 2D map, cf. Figure~\ref{fig:som}, which has been distorted in this 53-D space to represent the data in an optimized way.

\cite{PWLE05} applied a supervised classification algorithm known as a Support Vector Machine on 3200 Hipparcos variable stars. Such algorithm is a supervised technics,
so an already classified object set has to be identified and the algorithm has to be trained on this set. We took 18 classes. In total 51 features were used to classify the objects: the colors $B-V$ and $V-I$,  the skewness, the median subtracted 10-percentiles and forty bins from the Fourier envelope. The performance of the classification has been estimated to 62\%. A principal component analysis was performed to see if we could reduce the number of dimensions of the feature space, in fact it slightly degraded the performance of the classification. One probable cause of the low performance of this classification is the presence of poorly classified stars in the training set.

Here we represent the classification by a confusion matrix cf. Table~\ref{origohne}. So that the good and poor performances of the method for the variability types can be readily identified. It is extremely important to have such representation, which allows to clearly describe the method performance.

\begin{landscape}
\begin{table}
\setlength{\tabcolsep}{0.8mm}
\parbox[b]{22cm}{
\begin{center}
\begin{tabular}{lcccccccccccccccccc}
\cline{1-19}
 & \multicolumn{18}{c}{\textbf{true}} \\
\hline
\hline
\rule[3mm]{0mm}{1mm} \textbf{predicted} & ACV & ACYG & BCEP&  DCEP&  DSCT   & EA &  EB&  ELL&   EW&  GCAS&    I &   L &   M & RRAB  & RS & SPB & SRA & SRB \\             \cline{1-19}                                                            
ACV  &  42  & 2  & 6   &   & 1   &   & 6  & 9  & 1  & 15  & 2  & 1   &    &    &  & 36   &    & \\
ACYG   &   & 4   &    &    &    &   & 2   &    &   & 1  & 2  & 2   &    &    &    &    &    & \\
BCEP  &  2   &   & 3   &    &    &    &    &    &   & 1  & 1   &    &    &    &    &    &    & \\
DCEP   &    &    &   & 65   &   & 1  & 1  & 1   &    &   & 2  & 2   &    &    &    &   & 1   & \\
DSCT  &  1   &   & 2   &   & 35  & 1  & 5  & 1   &   & 1  & 1   &    &   & 2   &   & 1   &    & \\
EA  &  1   &    &    &    &  & 107  & 20   &    &   & 2  & 2   &    &    &    &    &    &    & \\
EB  &  2   &    &   & 1  & 1  & 12  & 60  & 2  & 11  & 11  & 4  & 2   &    &    &    &    &    & \\
ELL   &    &    &    &    &    &    &    &    &    &    &    &    &    &    &    &    &    & \\
EW   &    &    &    &   & 1   &   & 5   &   & 28   &    &    &    &    &    &    &    &    & \\
GCAS  &  5   &   & 1  & 1   &    &   & 2   &    &   & 32  & 3  & 2   &    &    &    &   & 1   & \\
I    &    &    &   & 3  & 1  & 4  & 4  & 3   &   & 5 & 124  & 71  & 1   &   & 1   &   & 1  & 11\\
L    &    &    &   & 2   &    &    &    &    &    &   & 16  & 34   &    &    &    &   & 6  & 16\\
M    &    &    &   & 1   &   & 1   &    &    &    &   & 1  & 1  & 63   &    &    &   & 5   & \\
RRAB   &    &    &    &    &    &    &    &    &    &    &    &    &   & 26   &    &    &    & \\
RS   &   & 1   &    &    &   & 2  & 1  & 1  & 1   &   & 6  & 5   &    &   & 20   &    &    & \\
SPB   &    &   & 1   &    &    &   & 2  & 1   &   & 1   &    &    &    &    &   & 1   &    & \\
SRA   &    &    &    &    &    &    &    &    &    &   & 1  & 2   &    &    &    &   & 1  & 2\\
SRB   &    &    &    &    &    &    &    &    &    &   & 4  & 10   &    &    &    &   & 3  & 23\\
\hline
\hline
\textbf{TP [\%]}  &   \textbf{79.2}  & \textbf{57.1}  & \textbf{23.1}  & \textbf{89.0}  & \textbf{89.7} &  \textbf{83.6} & \textbf{55.6}  &\textbf{0}   &\textbf{68.3}   &\textbf{46.4}  & \textbf{73.4} & \textbf{25.8} & \textbf{98.4}  & \textbf{92.9}  & \textbf{95.2}  &\textbf{2.6}  & \textbf{5.6} & \textbf{44.2} \\ 
\hline
\hline
objects/class  &   53  & 7  & 13  & 73  & 39 &  128 & 108  & 18   &41   &69  & 169 & 132 & 64  & 28  & 21  &38  & 18 & 52  \\
\hline
\end{tabular}
\caption{Confusion matrix for the classification results. We show the numbers of objects in the true versus the predicted classes. For better clarity, only non-zero entries are shown. The line \textbf{TP} presents the percentage of correctly identified stars. The last line shows the number of objects per class in the validation set. Summing up the numbers on the diagonal of the confusion matrix and dividing by the total number of objects in the validation set (1071) yields an estimate of the overall classification performance. In this case, we find 62.4~\% of correctly identified objects. From \cite{PWLE05}. \label{origohne}}
\end{center}
}
\end{table} 
\end{landscape}

\subsection{Visualization of the data and catalogue exploration}

Dr Laszlo Kiss once said that the IAU Coll. 193 in New Zealand on variable stars in 2003 was a celebration of {\sc ogle} data. This is thanks to the open policy of {\sc ogle} group who has made data publicly available for many years. {\sc Ogle} developed also an interface for the variable stars which allows to get access to any variable star they classified variable for the LMC and SMC, that are 68,000 stars \citep{KZetal01}. A light curve and an identification chart of every star can be displayed. This is indispensable to be able to browse through a catalogue, looking at individual cases. For large surveys, the possibility to be able to work on sub-samples is also crucial. It is permitting to get a good hint of the quality of a global and automated analysis.

Obviously there are other ways to display the data on a more global point of view, making 2D and 3D plots with each axis representing a given quantity.
However there are other ways, like Self-Organizing Maps, which can be used not only to classify, but also to visualize a data set, color scales, and contour plots are different ways to represent the data, as presented in Figure~\ref{fig:som}.

\cite{PPetal05} proposed to compute correlation coefficients of different folded curves for a given catalogue, so that to extract the least likely folded curves from the considered catalogue. In a way this is exploring the borders of a catalogue. He applied it to {\sc ogle} and {\sc macho} data for various groups like eclipsing binaries, RR Lyrae stars and Cepheids. The extracted cases can be spurious classifications, or peculiar possibly interesting behaviors.

\section{Global pipeline}

\begin{figure*}[!ht]
\plotone{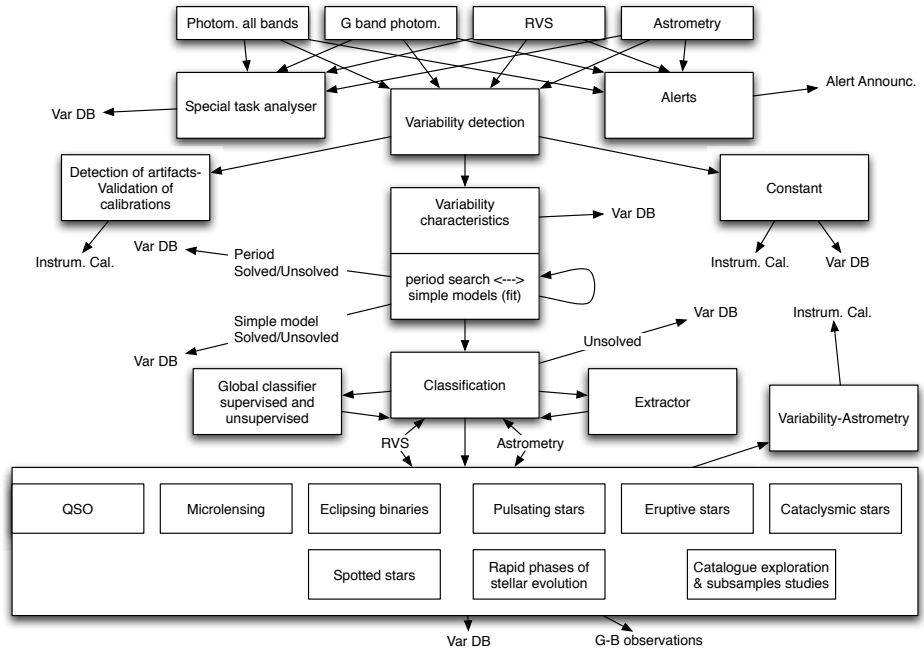}
\caption{Draft organigram of the variability analysis foreseen for Gaia.}
\label{fig:gvaran}
\end{figure*}

We continue the example of the Gaia mission, and present a possible scheme for a global analysis cf. Figure~\ref{fig:gvaran}. The number of sources is extremely large, and the procedures should be fully automated. The study should start as soon as calibrated data are available.
We find in the Figure~\ref{fig:gvaran} the different steps we mentioned in this article, the detection of variability, the characterization of the variability, period search, simple model fits (such as trend model, or Fourier series), classification (supervised or unsupervised), and finally specific studies on the classified groups. For example, it is envisioned to apply the Wilson-Devinney code for the stars classified as SB2 eclipsing binaries. 
At several steps, information from the variability analysis should be stored in a database (Var DB) and when a classification is determined with a estimated low error level, announcements should be done during the mission.

\section{Conclusion}

Surveys are producing a flood of data, either from ground based or space observations. The analysis of these large scale multi-epoch photometric surveys is a challenge. Thanks to these surveys, many new objects, phenomena are discovered and new research fields are created or made accessible.

In this current technologically driven research environment, there is a trade off between the spent efforts to maximize the return of a survey and to wait for the next generation surveys, which may make easier some aspects of the scientific exploitation, because of the improved precision and the larger number of objects for instances.
In this perspective some space missions are peculiar, because the time between consecutive generations of missions is very large. We could mention Hipparcos and Gaia, which will be separated by 12-14 years. 

There are also new ideas and improvements of software. Here we just mention one by  \cite{CARL98} and \cite{CA00}: Difference Image Analysis reaches high photometric precision in crowded fields.

The usage of certain modern Software is becoming more common, for example neural networks, genetic algorithms, self-organizing maps.

The goal is to maximize the scientific return of a survey with a reasonable effort. So efficient methods should be elaborated and compared; automated software pipelines designed and applied. These developments and choices are posing many questions, some answers are given, as presented in this article, but there is still a lot of margin in inventing novel methods, and optimizing procedures.

\end{document}